\documentclass[prd,onecolumn,superscriptaddress,showpacs,nofootinbib,preprintnumbers]{revtex4}

\usepackage{amsmath}
\usepackage{amsfonts}
\usepackage{graphicx}
\usepackage{dcolumn}
\usepackage{hyperref}

\def\R{{\cal R}}

\def\rd{{\rm d}}

\def\be{\begin{equation}}
\def\ee{\end{equation}}
\def\bea{\begin{eqnarray}}
\def\eea{\end{eqnarray}}

\def\GB{{\rm GB}}

\def\5{\overline 5}

\begin{document}

\title{String-inspired cosmology: Late time transition from scaling
matter era to dark energy universe caused by a Gauss-Bonnet coupling}

\author{Shinji Tsujikawa}
\email{shinji@nat.gunma-ct.ac.jp}
\affiliation{Department of Physics, Gunma National College of
Technology, Gunma 371-8530, Japan}

\author{M.~Sami}
\email{sami@iucaa.ernet.in}

 \affiliation{Department of physics,
Jamia Millia Islamia, New Delhi, India} \affiliation{Centre for
theoretical physics, Jamia Millia Islamia, New Delhi, India}

\date{\today}

\vskip 1pc
\begin{abstract}

The Gauss-Bonnet (GB) curvature invariant coupled to 
a scalar field $\phi$ can lead to an exit from a scaling 
matter-dominated epoch
to a late-time accelerated expansion, which is attractive
to alleviate the coincident problem of dark energy.
We derive the condition for the existence of cosmological
scaling solutions in the presence of the GB coupling
for a general scalar-field Lagrangian density $p(\phi, X)$,
where $X=-(1/2)(\nabla \phi)^2$ is a kinematic term of
the scalar field. The GB coupling and the
Lagrangian density are restricted to be in the form
$f(\phi) \propto e^{\lambda \phi}$ and
$p=Xg (Xe^{\lambda \phi})$, respectively, where
$\lambda$ is a constant and $g$ is an arbitrary function.
We also derive fixed points for such a scaling Lagrangian
with a GB coupling $f(\phi) \propto e^{\mu \phi}$ and clarify
the conditions under which the scaling matter era is followed by
a de-Sitter solution which can appear in the presence of
the GB coupling.
Among scaling models proposed in the current literature,
we find that the models which allow such a cosmological evolution are
an ordinary scalar field with an exponential potential and
a tachyon field with an inverse square potential,
although the latter requires a coupling between
dark energy and dark matter.

\end{abstract}


\maketitle
\vskip 1pc

\section{Introduction}

A great deal of efforts in modern cosmology, over the course of the
past decade, have been made in trying to understand the nature and
the origin of dark energy which is responsible for late-time
acceleration of the universe \cite{obser,review,CST}. Although
cosmological constant is the simplest candidate of dark energy
consistent with current observations, it suffers from a severe
fine-tuning problem from a field theoretic point of view
\cite{Weinberg}. It is then tempting to find alternative models of
dark energy which have dynamical nature unlike cosmological
constant. An accelerated expansion can be realized by using a scalar
field whose origin may be found in superstring or supergravity
theories. Thus modern particle physics has been an attractive
supplier for the construction of scalar-field dark energy
models--such as quintessence \cite{quin}, K-essence \cite{Kes},
tachyon \cite{tachyon}, dilatonic models \cite{Gas}, and ghost
condensate \cite{ghostcon,PT04}.

Although the dynamically evolving field models have an edge
over the cosmological constant scenario, they too, in general, involve
fine-tuning of initial values of the field and the parameters of the model.
Secondly, nucleosynthesis places stringent restriction on any additional
degree of freedom (over and above the particle physics standard
model degrees of freedom) which translates into a constraint on the ratio
of the field energy density to the energy density of the background
fluid \cite{CST}. These problems can be addressed by employing scalar field
models exhibiting {\it scaling solutions}
\cite{Fer,CLW,scaling,TS04,Tsuji06,AQTW}.
For instance, a standard scalar field $\phi$ with an exponential
potential  in General Relativity (GR) \cite{Fer,CLW}
can mimic the background (radiation/matter) remaining
sub-dominant so as to respect the nucleosynthesis constraint
\cite{Nelson}. The scaling solutions as dynamical
attractors can successfully address the second problem and can also considerably
alleviate the fine-tuning problem of initial conditions.

For a viable cosmological evolution, the scalar field should remain
unimportant for most of the cosmic evolution and should emerge from
the hiding epoch only at late times to give rise to the accelerated
expansion. Since scaling solutions are non-accelerating, the model
should be supplemented by an additional mechanism allowing the field
to exit from the scaling regime at late times. There are several
ways of achieving this goal:
\begin{itemize}
\item (i) At late times the potential becomes shallow to lead to the
accelerated expansion.
For example this is realized by the double exponential potential
given by
$V(\phi)=V_{0} (e^{-\lambda \kappa \phi}+e^{-\mu \kappa \phi})$
with $\lambda^2>20$ and $\mu^2<2$ \cite{Nelson} (here $\kappa^2=8\pi G$
with $G$ being gravitational constant).

\item (ii) A local minimum of the field potential appears at late
times. The examples of this type of models are
$V(\phi)=V_{0} [\cosh (\lambda \kappa \phi)-1]^n$ \cite{Sahni} and
$V(\phi)=V_{0} e^{-\lambda \kappa \phi}[A+(\kappa \phi -B)^2]$
\cite{Alb}.
In the former case the acceleration occurs for $n<1/2$ on average, whereas
in the latter case the system always approaches a de-Sitter universe
after a damped oscillation of the field.

\item (iii) Assisted  quintessence \cite{KLT}.
This is realized by considering many fields $\phi_i$ with
the sum of the exponential potential $V=\sum A_i
e^{-\lambda_i \kappa \phi_i}$, since the presence of
many scalar fields leads to the smaller effective coupling,
i.e., $\lambda_{\rm eff}^{-2}=
\sum \lambda_{i}^{-2}$ \cite{Liddle}.
\end{itemize}

Although these are interesting attempts to provide a successful exit
from the scaling regime, it is fair to say that the aforementioned
mechanisms are still phenomenological. It is of great importance to
find cosmological models from fundamental theories of particle
physics like string theory which may incorporate the findings of the
phenomenological studies. String theory necessarily includes scalar
fields (dilaton $\&$ modulus fields); the low energy string
effective action also gives rise to a modification of standard GR in
terms of Riemann invariants coupled to scalar field(s)
\cite{reviewstring}. The modifications are important to address the
problems of past \cite{past} and future \cite{future} singularities.
Out of these corrections, the Gauss-Bonnet (GB) term is of a special
relevance. The GB term is a topological invariant quantity which
contributes to the dynamics in 4 dimensions provided it is coupled
to a dynamically evolving scalar field. It is the unique invariant
for which the highest (second) derivative occurs linearly in the
equations of motion thereby ensuring the uniqueness of solutions.
Nojiri {\it et al.} \cite{NOS} studied scalar-Gauss-Bonnet cosmology
to demonstrate the existence of a particular dark energy solution.
Such a solution exists in the presence of an exponential potential
$V(\phi) \propto e^{-\lambda \kappa \phi}$ and the GB coupling
$f(\phi) \propto e^{\mu \kappa \phi}$. More recently Koivisto and
Mota \cite{KM06} showed that the GB coupling allows a possibility to
lead to the exit from the scaling matter era to the dark energy
dominated universe for the same model. See
Refs.~\cite{CTS,Neupane,NO05,Cog06,Cal,Sami06,Annalen,Sanyal,Andrew} for
other aspects of the GB dark energy scenario.

In this paper we shall clarify the form of the GB coupling $f(\phi)$ for the
existence of scaling solutions by starting from a very general
scalar-field Lagrangian density $p(\phi, X)$ where $X$ is a kinematic
term of the field. This includes a wide variety of scalar-field models
such as quintessence, K-essence, tachyon, and ghost condensate.
The demand for scaling solutions not
only determines the field Lagrangian to be $p(\phi,X)=Xg(Xe^{\lambda \phi})$
($g$ is an arbitrary function) but also fixes the GB coupling in the
form $f(\phi) \propto e^{\lambda \phi}$ (here we use the unit
$\kappa^2=1$).
Thus our results are useful to construct viable dark energy models which
possess scaling solutions in the matter-dominated epoch.
We then investigate the fixed points for the scaling Lagrangian
$p(\phi,X)=Xg(Xe^{\lambda \phi})$ with a GB coupling
$f(\phi) \propto e^{\mu \phi}$.
We show the existence of a pure de-Sitter fixed point which exists
only for specific models and study
a possibility to obtain a scaling matter era that finally approaches
the de-Sitter solution.
A standard field with an exponential potential is a specific model
to realize such a cosmological evolution provided that $\mu>\lambda$.
A tachyon field also possesses the de-Sitter fixed point, but it does
not have a scaling matter epoch. However the scaling matter era followed
by the de-Sitter solution is possible if dark energy couples to dark
matter. We shall discuss cosmological dynamics in such cases
in details.

\section{The condition for the existence of scaling solutions in Gauss-Bonnet cosmology}

The model we study is given by the following action
\begin{equation}
S=\int{\rm{d}}^{4}x\sqrt{-g_{M}}\left[\frac{1}{2\kappa^2}R
+p(X,\phi)-f(\phi)R_{\GB}^2 \right]+
S_{m}[\phi,\psi_{i},g_{\mu\nu}],\label{eqn:action}
\end{equation}
where $g_M$ is a metric determinant,
$R$ is a Ricci scalar,
$p(\phi, X)$ is a Lagrangian density in terms of a scalar field $\phi$
and a kinematic energy $X=-(1/2)g^{\mu\nu}\partial_{\mu}\phi\:
\partial_{\nu}\phi$.
The Gauss-Bonnet (GB) term,
$R_{\GB}^2 \equiv R^2-4R_{\mu\nu}R^{\mu\nu}+
R_{\alpha\beta\mu\nu}R^{\alpha\beta\mu\nu}$,
couples to the field $\phi$ through the coupling $f(\phi)$.
We allow for an arbitrary coupling between the matter fields
$\psi_{i}$ and the scalar field $\phi$.
We assume that the field $\phi$ is coupled to a barotropic
perfect fluid (energy density $\rho_{m}$
and pressure density $p_{m}$)
with a coupling given by \cite{Lucacoupled,LucaST,GNST}
\begin{equation}
Q=-\frac{1}{\rho_{m}\sqrt{-g_{M}}}
\frac{\delta S_{m}}{\delta \phi}\,.
\end{equation}
Note that the energy density of the scalar field is given by
$\rho=2X(\partial p/\partial X)-p$ with an equation
of state $w_{\phi}=p/\rho$.
In what follows we shall use the unit $\kappa^2=8\pi G=1$.

In the flat Friedmann-Robertson-Walker background with a scale
factor $a$, the equations for $\rho$ and $\rho_m$ are
\bea
\label{rho1}
& & \dot{\rho}+3H(1+w_\phi)\rho=-Q \rho_m \dot{\phi}
-24f'(\phi) \dot{\phi} H^2 (H^2+\dot{H})\,, \\
\label{rho2}
& & \dot{\rho}_m+3H(1+w_m)\rho_m=
+Q \rho_m \dot{\phi}\,,
\eea
where $H \equiv \dot{a}/a$ is a Hubble parameter.
A dot and a prime represent derivatives in terms of
cosmic time $t$ and the field $\phi$, respectively.
The Hubble parameter satisfies the constraint equation:
\bea
\label{Hubble}
3H^2=\rho_m+\rho+24f'(\phi) \dot{\phi} H^3\,.
\eea
Then the energy density of the GB term is given
by $\rho_{\GB}=24f'(\phi) \dot{\phi} H^3$.
We define the fractional densities of the field $\phi$,
the GB term, and the matter fluid, respectively, as
\bea
\Omega_{\phi} \equiv \frac{\rho}{3H^2}\,,\quad
\Omega_{\GB} \equiv 8f'(\phi) \dot{\phi}H\,, \quad
\Omega_{m} \equiv \frac{\rho_{m}}{3H^2}=
1-\Omega_{\phi}-\Omega_{\GB}\,.
\eea

Scaling solutions are characterised by constant values of
$\Omega_{\phi}$, $\Omega_{\GB}$ and $\Omega_{m}$,
or equivalently a constant ratio between each energy density.
Then this gives the condition:
\bea
\label{scacon}
\frac{\rd \ln \rho}{\rd N}=
\frac{\rd \ln \rho_{m}}{\rd N}=
\frac{\rd \ln \rho_{\GB}}{\rd N} \equiv
-3(1+w_{\rm eff})\,,
\eea
where $w_{\rm eff}$ is an effective equation of state
related to the Hubble rate via the relation
\bea
\label{dotH}
\frac{\dot{H}}{H^2}=-\frac32
(1+w_{\rm eff})\,.
\eea
We also note that $w_{\phi}$ and $w_{m}$ are assumed
to be constants in the scaling regime.
{}From Eqs.~(\ref{rho1}) and (\ref{rho2})
together with the scaling condition (\ref{scacon}), we find
\bea
\label{Qdphio}
Q \frac{\rd \phi}{\rd N}=
\frac{3(w_m-w_{\phi})\Omega_{\phi}
-\Omega_{\GB} (1+\dot{H}/H^2)}
{\Omega_{\phi}+\Omega_{m}}\,.
\eea
Substituting this relation for Eq.~(\ref{rho2})
and using the relations (\ref{scacon}) and
(\ref{dotH}), we obtain
\bea
w_{\rm eff}=\frac{2(w_m \Omega_m+w_{\phi}\Omega_{\phi}
-\Omega_{\GB}/6)}{1+\Omega_{\phi}+\Omega_{m}}\,.
\eea
Hence Eq.~(\ref{Qdphio}) can be rewritten as
\bea
\label{Qdphi}
Q \frac{\rd \phi}{\rd N}=
\frac{1}{\Omega_{\phi}+\Omega_{m}}
\left[ 3(w_m-w_{\phi})\Omega_\phi+
\Omega_{\GB}
\frac{\Omega_{\phi}(1+3w_{\phi})+\Omega_m (1+3w_{m})}
{1+\Omega_\phi+\Omega_m}
\right] \equiv C\, (={\rm const})\,.
\eea
From these equations and the definition of $X$, we find
\begin{equation}
2X=H^{2}\left(\frac{{\rm{d}}\phi}
{{{\rm{d}}N}}\right)^{2}\;\;\propto\;\;
\frac{\rho}{{Q^{2}}}\;\;\propto\;\;\frac{p(X,\phi)}{Q^{2}}\,,
\label{eqn:X-proportional-to}
\end{equation}
which gives
\begin{equation}
    \frac{{\rm{d}}\ln X}{{\textrm{d}}N}=-
    3(1+w_{\rm eff})-2\frac{{\rm{d}}\ln
    Q}{{\rm{d}}N}\,.\label{eqn:dlnX-dN}
\end{equation}

{}From Eq.~(\ref{scacon}) together with $p \propto \rho$,
the Lagrangian density $p=p(\phi, X)$
satisfies the differential equation
\begin{equation}
\label{pdef}
\frac{\partial \ln p}{\partial X} \frac{\rd X}
{\rd N}+\frac{\partial \ln p}{\partial \phi}
\frac{\rd \phi}{\rd N}=-3(1+w_{\rm eff})\,.
\end{equation}
Substituting Eqs.~(\ref{Qdphi}) and (\ref{eqn:dlnX-dN})
for Eq.~(\ref{pdef}), we arrive at
\begin{equation}
\label{master}
\left[1+\frac{2}{\lambda Q}
\frac{\rd Q(\phi)}{\rd \phi}\right]\frac{\partial \ln
p}{\partial \ln X}-\frac{1}{\lambda}
\frac{\partial \ln p}{\partial \phi}=1\,,
\end{equation}
where
\begin{equation}
\label{lamdef}
\lambda (\phi) \equiv \frac{3(1+w_{\rm eff})}{C}Q(\phi)=
\frac{2(1-\Omega_{\GB})[3(1+w_m \Omega_m +w_{\phi}
\Omega_{\phi})-2\Omega_{\GB}]}
{3(w_{m}-w_{\phi})\Omega_{\phi} (2-\Omega_{\GB})
+\Omega_{\GB}
[1-\Omega_{\GB}+3(w_m \Omega_m +
w_{\phi}\Omega_{\phi})]}Q(\phi)\,.
\end{equation}
Equation (\ref{master}) is the same form of equation which
was obtained in the absence of the GB coupling \cite{AQTW}.
Following the procedure in Ref.~\cite{AQTW},
Eq.~(\ref{master}) restricts the form of the Lagrangian
density to be
\begin{equation}
\label{gene}
p=X Q^2(\phi) g \left[X Q^2(\phi) \exp
\left( \int^{\phi} \lambda(\varphi)
\rd \varphi \right)\right]\,,
\end{equation}
where $g$ is an arbitrary function.
If $Q$ is a constant, this reduces to a simple form:
\begin{equation}
\label{scalag}
p=X g (Y)\,,\quad
Y \equiv X e^{\lambda \phi}\,,
\end{equation}
where $p$ is redefined so that a constant factor is
removed from Eq.~(\ref{gene}).
This form of the scaling Lagrangian was first derived
in Ref.~\cite{PT04} in the absence of the GB coupling
(see also Ref.~\cite{TS04} for the extension of the analysis
to the case $H^2 \propto \rho^n$).
Even if the GB term is present, we have shown that
the Lagrangian takes the same form with a modified
value of $\lambda$ given in Eq.~(\ref{lamdef}).
For a later convenience we present several models which belong to
the scaling Lagrangian (\ref{scalag}).

\begin{itemize}
\item (i) An ordinary field
with an exponential potential, i.e., $p=X-ce^{-\lambda \phi}$ \cite{CLW}.
This corresponds to the choice
\begin{equation}
\label{mo1}
g(Y)=1-c/Y\,.
\end{equation}

\item (ii) A dilatonic ghost condensate model, i.e.,
$p=-X+ce^{\lambda \phi}X^2$  \cite{PT04}.
This corresponds to the choice
\begin{equation}
\label{mo2}
g(Y)=-1+cY\,.
\end{equation}

\item (iii) A tachyon field, i.e.,
$p=-V(\varphi)\sqrt{1-\dot{\varphi}^2}$ with
$V(\varphi) \propto \varphi^{-2}$ \cite{Abramo,Ruth,CGST}.
This corresponds to the choice
\begin{equation}
\label{mo3}
g(Y)=-(c/Y)\sqrt{1-2Y}\,.
\end{equation}
Note that the  field $\varphi$ is related to $\phi$
via the relation $\varphi=(2/\lambda) e^{\lambda \phi}/2$.

\end{itemize}

{}From Eq.~(\ref{Qdphi}) we find $\dot{\phi}/H=C/Q$.
Since the GB energy density, $\rho_{\GB}=24f'(\phi)
\dot{\phi}H^3$, satisfies the relation
$\rd \ln \rho_{\GB}/\rd N=-3(1+w_{\rm eff})$
for scaling solutions, one gets
\begin{equation}
\frac{f''(\phi)}{f'(\phi)} \frac{\dot{\phi}}{H}=3(1+w_{\rm eff})\,.
\end{equation}
This yields $f''(\phi)/f'(\phi)=3(1+w_{\rm eff})
Q/C=\lambda$, which gives
\begin{equation}
\label{fform}
f'(\phi)=\alpha e^{\lambda \phi}\,,
\end{equation}
where $\alpha$ is an integration constant.
This means that the GB coupling is restricted to be in the form
$f(\phi) \propto e^{\lambda \phi}$ for the existence
of scaling solutions.

In the case of a non-relativistic dark matter ($w_m=0$),
the quantity $C$ is given by
\begin{equation}
C=\frac{1-\Omega_{\phi}-\Omega_m-
6 w_{\phi} \Omega_{\phi}}
{1+\Omega_{\phi}+\Omega_{m}}\,,
\end{equation}
whereas the effective equation of state is
\begin{equation}
w_{\rm eff}=-\frac{1-\Omega_{\phi}-\Omega_m-
6 w_{\phi} \Omega_{\phi}}
{3(1+\Omega_{\phi}+\Omega_{m})}\,.
\end{equation}
Hence one has a simple relation $w_{\rm eff}=-C/3$.
{}From Eq.~(\ref{lamdef}) we get $C=3Q/(Q+\lambda)$.
Then from Eq.~(\ref{Qdphi}) we find the following
relation along the scaling solution:
\begin{equation}
\label{dphiHre}
\frac{\dot{\phi}}{H}=\frac{3}{Q+\lambda}\,.
\end{equation}
This is the same relation as in the case where the
GB term is absent \cite{PT04,TS04}.

\section{Autonomous equations and fixed points}

In this section we shall obtain fixed points for the scaling
Lagrangian (\ref{scalag}).
We take into account the contribution of both non-relativistic
dark matter ($w_{m}=0$) and radiation ($w_r=1/3$),
in which case the Friedmann equation is
\begin{equation}
\label{H2}
3H^2=\rho_m+\rho_r+\rho+24f'(\phi)\dot{\phi}H^3\,.
\end{equation}
Here the energy density of the scalar field is given by
$\rho=X(g+g_1)$, where $g_n \equiv Y^n g^{(n)}(Y)$
with $g^{(n)}(Y)$ being the $n$-th derivative in terms of $Y$.
We assume that the field $\phi$ is coupled to dark matter
with a coupling $Q$ and that it is uncoupled to radiation,
as it is the case in the context of scalar-tensor
theories \cite{LucaST}.
Hence the equations for the energy densities of dark matter
and radiation are
\bea
\label{rhom}
& &\dot{\rho}_{m}+3H \rho_m=+Q\rho_m \dot{\phi}\,, \\
\label{rhor}
& & \dot{\rho}_{r}+4H\rho_r=0\,.
\eea

The scalar-field equation (\ref{rho1}) can be written as
\bea
\label{ddotphi}
\ddot{\phi}+3AH (g+g_1) \dot{\phi}+\lambda X
\left[ 1-A(g+2g_1) \right]+24f'(\phi)A H^2
(H^2+\dot{H})+AQ\rho_m=0\,,
\eea
where $A$ is defined by
\bea
\label{Adef}
A \equiv (g+5g_1+2g_2)^{-1}\,.
\eea

We shall define the following dimensionless quantities:
\bea
\label{xidef}
x_1 \equiv \frac{\dot{\phi}}{\sqrt{6}H}\,, \quad
x_2 \equiv \frac{e^{-\lambda \phi/2}}{\sqrt{3}H}\,, \quad
x_3 \equiv f'(\phi)H^2\,, \quad
x_4 \equiv \frac{\sqrt{\rho_r}}{\sqrt{3}H}\,.
\eea
Then the energy fractions are
\bea
\label{Omedef}
\Omega_{\phi} = x_{1}^2 (g+2g_1)\,,\quad
\Omega_{\GB}=8\sqrt{6} x_1 x_3\,, \quad
\Omega_{r} =x_4^2\,, \quad
\Omega_{m}=
1-\Omega_{\phi}-\Omega_{\GB}-\Omega_r\,.
\eea
The variable $Y$ defined in Eq.~(\ref{scalag})
can be expressed as
\bea
Y=x_1^2/x_2^2\,.
\eea
The GB coupling $f'(\phi)$ takes the form (\ref{fform})
for the existence of scaling solutions.
We shall derive autonomous equations
for a more general case:
\bea
\label{fdphidef}
f'(\phi)=\alpha e^{\mu \phi}\,.
\eea
Since $\mu=\lambda$ for scaling solutions,
one has the relation
\bea
\label{x2x3re}
x_2^2 x_3=\alpha/3\,.
\eea
In this case the variable $x_3$ is not needed
to study a dynamical system.

Taking the time-derivative of Eq.~(\ref{H2})
with the use of Eqs.~(\ref{rhom}) and (\ref{rhor}),
we obtain
\bea
\label{dotH2}
2\dot{H}=-2X(g+g_1)-\rho_m-\frac43 \rho_r
+8\left[ f''(\phi) \dot{\phi}^2 H^2+f'(\phi) H
(\ddot{\phi}H+2\dot{\phi}\dot{H}-\dot{\phi}H^2)
\right]\,.
\eea
Eliminating the $\ddot{\phi}$ term from 
Eqs.~(\ref{ddotphi}) and (\ref{dotH2}),
one gets the following relation
\bea
\label{dotH3}
(1-8\sqrt{6}x_1x_3+96A x_3^2)
\frac{\dot{H}}{H^2} &=& -\frac12 (3+3gx_1^2+x_4^2)
\nonumber \\
&&+4x_3\biggl[ 2\sqrt{6}x_1+3(2\mu-\lambda) x_1^2+
3A \biggl\{-\sqrt{6} (g+g_1)x_1 +(Q+\lambda)x_1^2
(g+2g_1) \nonumber \\
&&-8x_3+Q(x_4^2-1)+8\sqrt{6}Qx_1 x_3
\biggr\}\biggr]\,.
\eea
The autonomous equations are
\bea
\label{auto1}
\frac{\rd x_1}{\rd N}&=&-3A(g+g_1)x_1+\frac{\sqrt{6}}{2}
\left[ A(Q+\lambda) (g+2g_1)x_1^2-\lambda x_1^2
+QA (x_4^2-1+8\sqrt{6} x_1 x_3 )\right] \nonumber \\
& &-4\sqrt{6}Ax_3
-(x_1+4\sqrt{6} Ax_3) \frac{\dot{H}}{H^2}\,,\\
\label{auto2}
\frac{\rd x_2}{\rd N}&=&-x_2\left(\frac{\sqrt{6}}{2}
\lambda x_1+\frac{\dot{H}}{H^2}\right)\,,\\
\label{auto3}
\frac{\rd x_3}{\rd N}&=& 2x_3\left(
\frac{\sqrt{6}}{2}\mu x_1+\frac{\dot{H}}{H^2}\right)\,,\\
\label{auto4}
\frac{\rd x_4}{\rd N}&=& -x_4\left(
2+\frac{\dot{H}}{H^2}\right)\,,
\eea
where $\dot{H}/H^2$ is given in Eq.~(\ref{dotH3}).

In what follows we will obtain fixed points in the absence of radiation
($x_4=0$). Let us consider the cases: (i) $\lambda=\mu$ and
(ii) $\lambda \neq \mu$ separately.
We shall study the case $\lambda>0$ without loss of generality.

\subsection{Case of $\lambda=\mu$}

As we showed in the previous section, there exist 
scaling solutions for $\lambda=\mu$.
Taking note of the relation (\ref{x2x3re}), neither $x_2$
nor $x_3$ is identical to zero if $\alpha  \neq 0$.
When $\alpha$ is negligibly small, $x_2$ and $x_3$
can be very small as well. In this case we may regard
$x_2 \approx 0$ or $x_3 \approx 0$ as approximate fixed points
in Eqs.~(\ref{auto2}) and (\ref{auto3}).
Here we do not consider such fixed points, but
in the next subsection we will discuss those cases 
when $\lambda$ does not equal to $\mu$.

Then from Eqs.~(\ref{auto2}) and (\ref{auto3}), the critical points
corresponding to $x_2 \neq 0$ and $x_{3} \neq 0$ satisfy
\bea
\label{re1}
\frac{\dot{H}}{H^2}=-\frac{\sqrt{6}}{2} \lambda x_1\,.
\eea
Then substituting Eq.~(\ref{re1}) for Eq.~(\ref{auto1}),
we get
\bea
\label{re2}
(Q+\lambda)(g+2g_1)x_1^2-Q-\sqrt{6}(g+g_1)x_1
+4x_3 \left[ \sqrt{6} (2Q+\lambda) x_1-2 \right]=0\,.
\eea
{}From Eqs.~(\ref{dotH3}) and (\ref{re1}), we find
\bea
\label{re3}
3gx_1^2-\sqrt{6} \lambda x_1+3=
8x_1 x_3 (2\sqrt{6}-3\lambda x_1)\,.
\eea
We note that the following useful relations hold
for the scaling Lagrangian (\ref{scalag}):
\bea
w_\phi=\frac{g}{g+2g_1}\,, \quad
w_{\phi}\Omega_{\phi}=gx^2\,, \quad
g+g_1=\frac{(1+w_{\phi})\Omega_{\phi}}{2x^2}\,.
\eea
Substituting these expressions for
Eqs.~(\ref{re2}) and (\ref{re3}) and eliminating
the $w_{\phi}\Omega_{\phi}$ term, we obtain
\bea
\left[2(Q+\lambda)x_1-\sqrt{6} \right]
(1-\Omega_{\phi}-\Omega_{\GB})=0\,.
\eea
This gives the following fixed points:
\begin{itemize}
\item
(a) Scaling solution:
$x_1=\frac{\sqrt{6}}{2(Q+\lambda)}$\,,
\item
(b) Scalar-field $\&$  GB dominated point:
$\Omega_{\phi}+\Omega_{\GB}=1$\,.
\end{itemize}

Since both points satisfy the relation (\ref{re1}),
the effective equation of state is
\bea
\label{weff}
w_{\rm eff}=-1+\frac{\sqrt{6}}{3}\lambda x_1\,.
\eea
In what follows we shall discuss two fixed points
separately.

\subsubsection{Scaling solutions}

First we recall that Eq.~(\ref{dphiHre}) also gives
$x_1=\frac{\sqrt{6}}{2(Q+\lambda)}$.
{}From Eq.~(\ref{weff}) the effective equation of
state is given by
\bea
w_{\rm eff}=-\frac{Q}{Q+\lambda}\,,
\eea
which is independent of the form of $p(\phi, X)$.
The accelerated expansion occurs for $w_{\rm eff}<-1/3$, i.e.,
$Q>\lambda/2$ or $Q<-\lambda$.

{}From Eq.~(\ref{re3}) we get
\bea
\label{x3}
x_{3}=\frac{2Q(Q+\lambda)+3g(Y)}
{8(4Q+\lambda)}\,,
\eea
where
\bea
\label{Yeq}
Y=\frac{x_1^2}{x_2^2}=\frac{9}{2(Q+\lambda)^2
\alpha}x_{3}\,.
\eea
Once the form of $g(Y)$ is specified, one gets
$x_3$ and $Y$ from Eqs.~(\ref{x3}) and
(\ref{Yeq}). This then determines
\bea
\Omega_{\phi}=\frac{3}{2(Q+\lambda)^2}
(g+2g_1)\,,\quad
\Omega_{\GB}=\frac{24}{Q+\lambda}x_3\,.
\eea

Let us consider an ordinary scalar field with an exponential
potential, i.e., the model given in Eq.~(\ref{mo1}).
In this case we get
\bea
\label{x3de}
x_3=\frac{6Q(Q+\lambda)+9 -
\sqrt{[6Q(Q+\lambda)+9]^2-192c\alpha (4Q+\lambda)
(Q+\lambda)^2}}{48(4Q+\lambda)}\,.
\eea
We have chosen a minus sign in Eq.~(\ref{x3de})
to recover $x_{3} \to 0$ as $\alpha \to 0$.
When $Q=0$ this reduces
\bea
x_3=\frac{3}{16\lambda}
\left[ 1-\sqrt{1-\frac{64}{27}c\alpha \lambda^3}
\right]\,,
\eea
and also we obtain
\bea
\Omega_{\phi}=\frac{3}{2\lambda^2}
\left[ 1+\frac{32c\alpha \lambda^3}
{27(1-\sqrt{1-64c\alpha \lambda^3/27})} \right]\,,
\quad
\Omega_{\GB}=\frac{9}{2\lambda^2}
\left[1-\sqrt{1-\frac{64}{27}c\alpha \lambda^3}
\right]\,.
\eea
If $c\alpha \lambda^3$ is much smaller than unity,
we find the following approximate fixed points:
\bea
x_2 \simeq \sqrt{\frac{3}{2c\lambda^2}}\,,
\quad
x_3 \simeq \frac29 c\alpha \lambda^2\,,
\label{x3ap}
\eea
together with
\bea
\Omega_{\phi} \simeq \frac{3}{\lambda^2}
\left(1-\frac{8}{27}c \alpha \lambda^3 \right)\,,
\quad
\Omega_{\GB} \simeq \frac{16}{3}c
\alpha \lambda\,,
\eea
which, in the limit $\alpha \to 0$, recovers the
scaling solution derived in Ref.~\cite{CLW,GNST}.

The above scaling solution is attractive to be used in
a matter era to alleviate the coincident problem, but it
does not give way to a late-time acceleration
since the scaling solution is a global
attractor \cite{Lucacoupled,Tsuji06, AQTW}.
If $\mu>\lambda$, however, the presence of the GB term
can lead to the exit from the scaling matter era, as we
will discuss later.

\subsubsection{Scalar-field $\&$ GB dominated point}

Since $\Omega_{\phi}+\Omega_{\GB}=1$ in this case,
this gives
\bea
\label{scalar1}
x_{1}^2 (g+2g_1)+8\sqrt{6}x_{1}x_{3}=1\,.
\eea
Making use of Eq.~(\ref{re2}) together with this equation,
we obtain
\bea
\label{scalar2}
16x_3=\lambda [x_1^2 (g+2g_1)+1]
-2\sqrt{6} (g+g_1)x_1\,.
\eea
If we specify the form of $g(Y)$, one can get $x_1$ and $x_3$
from the above equations.
Let us consider the standard field with an exponential potential,
i.e., $p=1-c/Y$. Then from Eq.~(\ref{scalar2}) we get
\bea
x_3=\frac{\lambda-\sqrt{6}x_1}{4(2+\sqrt{6}\lambda x_1)}\,.
\eea
Using Eq.~(\ref{scalar1}) one obtains the following equation
for $x_1$:
\bea
\label{defeq}
3(\lambda- \sqrt{6} x_1)(\sqrt{6}\lambda x_1^3-10x_1^2
+\sqrt{6}\lambda x_1-2)+4c\alpha (2+\sqrt{6} \lambda
x_1)^2=0\,.
\eea
When $\alpha=0$ this has a solution $x_1=\lambda/\sqrt{6}$,
which can lead to an accelerated expansion for $\lambda<\sqrt{2}$.
Let us obtain the solution for (\ref{defeq})
perturbatively under the assumption that $\alpha$ is much smaller
than unity.
Substituting $x_1=\lambda/\sqrt{6}+\epsilon$ for Eq.~(\ref{defeq}),
we find the following approximate relation
\bea
x_{1}=\frac{\lambda}{\sqrt{6}}-
\frac{4\sqrt{6} c\alpha (\lambda^2+2)}
{3(6-\lambda^2)}\,, \quad
x_{2}=\sqrt{\frac{1}{c} \left(1-\frac{\lambda^2}{6}\right)}\,, \quad
x_{3}=\frac{2c\alpha}{6-\lambda^2}\,,
\eea
and
\bea
\label{OmeGB}
\Omega_{\GB}=\frac{16c\alpha \lambda}{6-\lambda^2}\,,
\quad
\Omega_{\phi}=1-\Omega_{\GB}\,.
\eea

Note that the effective equation of state is given by
\bea
\label{w}
w_{\rm eff}=-1+\frac{\lambda^2}{3}
-\frac{8c\alpha \lambda (\lambda^2+2)}{3(6-\lambda^2)}
=-1+\frac{\lambda^2}{3}-\frac{\lambda^2+2}{6}\Omega_{\GB}\,.
\eea
Thus the contribution of the GB term tends to reduce $w_{\rm eff}$.
In the absence of the GB coupling the late-time acceleration occurs
for $\lambda^2<2$.
In this case the scaling matter era is not present, since the
existence of it requires the condition $\lambda^2>3$ \cite{CLW}.
Hence this case is not attractive to solve the coincident problem,
although the standard matter-dominated epoch is followed by
an accelerated expansion due to a shallow potential as in the case
of a cosmological constant.
When the GB term is present, there is a contribution of its energy
density to the late-time acceleration.
However, as we will see in a later section, the contribution of the
GB term is restricted to be small from the constraint that it does
not disturb the cosmological evolution
during radiation and matter eras.

\subsection{Case of $\lambda \neq \mu$: Exit from the scaling regime}

Strictly speaking, scaling solutions are present
only for $\lambda=\mu$ in the presence of the
GB coupling.
Even when $\lambda \neq \mu$, however,
they can exist approximately as long as the contribution of the
GB term is negligibly small.
This corresponds to a situation in which $x_3$ is
very much smaller than 1.
Note that $x_3$ can not be exactly zero since
the relation between $x_2$ and $x_3$ is
\bea
\label{re2d}
x_2^2 x_3=\frac{\alpha}{3} e^{(\mu-\lambda)\phi}\,.
\eea
Still one can regard $x_{3} \approx 0$ as an approximate fixed
point satisfying this relation.
Then we approximately obtain the following 4 fixed points
(A), (B), (C) and (D)
using the results of Refs.~\cite{Tsuji06,AQTW}
recently obtained for $x_3=0$.
Hereafter, when we write $x_2=0$ or $x_3=0$,
it means that they are not exactly zero but are
very small values.

\subsubsection{$x_3 = 0$ {\rm and}
$\sqrt{6}\lambda x_1/2=-\dot{H}/H^2$}

\begin{itemize}
\item (A) Scaling solutions.

They are characterized by
\bea
\label{scalmune}
x_1=\frac{\sqrt{6}}{2(Q+\lambda)}\,,\quad
w_{\rm eff}=-\frac{Q}{Q+\lambda}\,, \quad
\Omega_{\phi}=\frac{Q(Q+\lambda)+3p_{,X}}{(Q+\lambda)^2}\,,
\quad
g(Y)=-\frac23 Q (Q+\lambda)\,,
\eea
where $p_{,X} \equiv \partial p/\partial X$.
Note that $x_1$ and $w_{\rm eff}$ are the same as in the case where
the GB term is present.
This solution is stable when the following conditions 
are satisfied \cite{AQTW}:
\bea
-\frac{Q}{Q+\lambda} \le \Omega_{\phi}<1\,, \quad A>0\,,
\eea
where $A$ is defined in Eq.~(\ref{Adef}).
The latter condition automatically holds when we impose the stability
of quantum fluctuations of the scalar field \cite{PT04}.
When $Q=0$ the scaling solution is stable for
$0 \le \Omega_{\phi} <1$, as required for the existence of itself.

\item (B) Scalar-field dominated solutions.

They are characterized by
\bea
\Omega_{\phi}=1\,,\quad
w_{\rm eff}=w_{\phi}=-1+\frac{\sqrt{6}\lambda}{3}x_1\,.
\eea
When $Q$ is positive, this solution is stable for
$x_1<\sqrt{6}/[2(Q+\lambda)]$.

\subsubsection{$x_3 = 0$ {\rm and}
$x_2 = 0$}

The solutions which satisfy $x_3=0$ and $x_2=0$ correspond
to kinematic fixed points. These exist only when
$g(Y=x_1^2/x_2^2)$ is non-singular,
namely, when $g$ is expanded as
\bea
\label{lagexist}
g=c_0-\sum_{n>0} c_{n} Y^{-n}\,,
\eea
where $c_0$ and $c_n$ are constants.
Since $g_n (x_ 2 \to 0)=0$ in this case, one can easily get
the following points from Eq.~(\ref{auto1}).

\item (C) $\phi$MDE solution.

This is characterized by
\bea
\label{phiMDE}
x_1=-\frac{\sqrt{6}Q}{3c_0}\,, \quad
x_2 = 0\,, \quad
\Omega_{\phi}=w_{\rm eff}=
\frac{2Q^2}{3c_0}\,.
\eea
When $c_{0}>0$, i.e., corresponding
to a non-phantom field, this solution is a saddle point.
Note that when $Q=0$ the $\phi$MDE is equivalent
to a standard matter-dominated epoch.

\item (D) Pure kinetic solutions.

These solutions are
\bea
x_1=\pm 1/\sqrt{c_0}\,, \quad
x_2 = 0\,, \quad
\Omega_{\phi}=w_{\rm eff}=1\,.
\eea
We need positive $c_{0}$ for their existence.
These are saddle or unstable nodes if $Q>0$.
Since $\Omega_{\phi}=w_{\rm eff}=1$, one can
use pure kinetic solutions neither for matter/radiation
eras nor for dark energy eras.

\end{itemize}

{}From Eq.~(\ref{auto2}) and (\ref{auto3}) we find that  
there are two other cases.

\subsubsection{$\sqrt{6}\mu x_1/2=-\dot{H}/H^2$ {\rm and} $x_2=0$}

\begin{itemize}

\item (E) Kinetic and GB dominated solutions.

Since $x_2=0$ in this case, the form of the function
$g(Y)$ is restricted to be in the form (\ref{lagexist}).
When $Q=0$ and $c_0=0$ these solutions satisfy
\bea
\label{Fx1}
F(x_1) \equiv
6\mu^2 x_1^4 -(24+\sqrt{6})\mu  x_1^3+[6(\mu^2+5)
4\sqrt{6}] x_1^2-5\sqrt{6} \mu x_1+6=0\,,\quad
x_3=\frac{\sqrt{6}x_1(2-\mu x_1)}{8(3\mu x_1-\sqrt{6})}\,,
\eea
together with an effective equation of state:
\bea
\label{weffki}
w_{\rm eff}=-1+\frac{\sqrt{6}}{3}
\mu x_{1}\,.
\eea
We have $w_{\rm eff}=-1/3$ for $x_1=\sqrt{6}/(3\mu)$.
Equation (\ref{Fx1}) possesses two real solutions.
If $\mu=10$, for example, we get
$(x_1, x_3, w_{\rm eff})=(0.08235, 1.40993, -0.3276),
(0.12454, 0.02236, 0.01689)$.
In both cases we do not have an accelerated expansion
($w_{\rm eff}<-1/3$).
When $\mu>0$ we find that the values of $x_1$ corresponding
to two real solutions of Eq.~(\ref{Fx1}) are larger
than $\sqrt{6}/(3\mu)$, which means $w_{\rm eff}>-1/3$ from
Eq.~(\ref{weffki}).
Hence one can not use these solutions for dark energy.

These fixed points correspond to the absence of the field potential,
in which case the accelerated solution has not
been found in Ref.~\cite{CTS}.
This is consistent with the above result.

\end{itemize}

\subsubsection{$\sqrt{6}\mu x_1/2=-\dot{H}/H^2$
{\rm and} $\sqrt{6}\lambda x_1/2=-\dot{H}/H^2$}

\begin{itemize}

\item{(F) de-Sitter point.}

\end{itemize}

When $\lambda \neq \mu$ there exists a fixed point $x_1=0$
from Eqs.~(\ref{auto2}) and (\ref{auto3}).
Since $\dot{H}/H^2=0$ in this case,
this corresponds to a de-Sitter solution.
In fact one has $w_{\rm eff}=-1$ from Eq.~(\ref{weff}).
Equation (\ref{dotH3}) gives
\bea
-\frac12 (3+3gx_1^2)+12Ax_{3}
\left[ -\sqrt{6} (g+g_1) x_1+(Q+\lambda) (g+2g_1)
x_1^2-8x_3-Q \right]=0\,.
\eea
Meanwhile Eq.~(\ref{auto1}) leads to
\bea
\label{Are}
A \left[ -\sqrt{6} (g+g_1) x_1+(Q+\lambda) (g+2g_1)x_1^2
-8x_3-Q \right]=0\,.
\eea
Then we find that the fixed point satisfies
\bea
\label{gx1}
gx_1^2=-1\,,\quad
x_1=0\,.
\eea
Once the form of $g(Y)$ is specified, one can get
$x_2$ by solving this equation.
{}From Eq.~(\ref{Are}) we get
\bea
x_3=\frac18 \left[ (Q+\lambda)(g+2g_1)x_1^2-Q
-\sqrt{6}(g+g_1)x_1 \right]\,.
\eea

Let us consider a standard scalar field with an exponential
potential, i.e., the model given by (\ref{mo1}).
Then the above de-Sitter solution corresponds to
\bea
\label{desitter0}
(x_1, x_2, x_3)=(0, 1/\sqrt{c}, \lambda/8)\,.
\eea
More generally Eq.~(\ref{gx1}) can be satisfied when $g=g(Y)$
is written in the form
\bea
\label{model2}
g(Y)=\left[c_{0}-(c/Y) \right]f(Y)\,,
\eea
where $f(Y)$ is a function which approaches a constant $d$
as $x_1 \to 0$.
In this case we obtain the following fixed point:
\bea
\label{desitter}
(x_1, x_2, x_3)=(0, 1/\sqrt{cd}, \lambda/8)\,.
\eea
The model (\ref{model2}) includes a tachyon field
with a Lagrangian given in Eq.~(\ref{mo3}).

In order for the existence of the de-Sitter solution, it is crucial to
have the term $c/Y$ in the expression of $g(Y)$.
For example, when $g(Y)$ is written as a sum of positive
powers of $Y^n$ such as a dilatonic ghost condensate model
given in Eq.~(\ref{mo2}), we do not have such a de-Sitter solution.

{}From Eqs.~(\ref{Omedef}) and (\ref{desitter}) we find that this
fixed point satisfies
\bea
\Omega_{\phi}=1\,, \quad
\Omega_{\GB}=0\,.
\eea
Since the field is frozen at the de-Sitter point, the GB term
does not have any contribution to the energy density of
the universe.

The appearance of the de-Sitter point comes from the presence
of the GB term. For example when $g(Y)=1-c/Y$,
Eq.~(\ref{ddotphi}) reduces to
\bea
\ddot{\phi}+3H\dot{\phi}-c\lambda  e^{-\lambda \phi}
+24f'(\phi)(H^2+\dot{H})=0\,,
\eea
where we dropped the coupling $Q$.
Since $f'(\phi)H_c^2=\lambda/8$ and $\dot{H}_c=0$
at the critical point, the effective potential
$V_{\rm eff}(\phi)$ for the field $\phi$ satisfies
the relation
\bea
\frac{{\rm d}V_{\rm eff}}{{\rm d}\phi}
=-c\lambda e^{-\lambda \phi}+3\lambda H_{c}^2\,.
\eea
The last term appears because of the presence of the GB term,
which gives rise to a potential minimum for the field $\phi$.
Then after the field drops down at this minimum,
the universe exhibits a de-Sitter expansion.

Let us consider the stability of the de-Sitter solution.
By perturbating Eqs.~(\ref{auto1}), (\ref{auto2})
and (\ref{auto3}) about the fixed point,
we obtain a $3\times 3$ Jacobian matrix ${\cal M}$
for perturbations $\delta x_1$, $\delta x_2$ and
$\delta x_3$ \cite{CST}
(we neglect the contribution of radiation).
For the model (\ref{mo1}) the eigenvalues of
the matrix ${\cal M}$ are
\bea
\label{eigen}
\lambda_1=-3, \quad
\lambda_{2, 3}=\frac32 \left[ -1 \pm
\sqrt{1+\frac{8\lambda (\lambda -\mu)}
{3(2+3\lambda^2)}} \right]\,.
\eea
This explicitly shows the following property for the
stability of the de-Sitter point:
\begin{itemize}
\item (i) Stable for $\mu>\lambda$.
\item (ii) Saddle for $\mu<\lambda$.
\end{itemize}
When $\mu=\lambda$ one has $\lambda_1=-3, \lambda_2=0$ and
$\lambda_3=-3$, which means that the de-Sitter solution
is marginally stable.

The same eigenvalues as given in Eq.~(\ref{eigen}) can be obtained
by demanding the conditions $\Omega_{\phi}=1$, $(g+g_1)x_1=0$ and
$A=1$ at the fixed point without specifying the form of $g(Y)$.
This includes the tachyon model (\ref{mo3}) with a choice $c=1$.
Provided that $\mu>\lambda$ the system falls down to the stable
de-Sitter point.

\section{Cosmological dynamics}

In this section we shall study cosmological dynamics
for the scaling Lagrangian (\ref{scalag}) with the 
GB coupling given in Eq.~(\ref{fdphidef}).
Our interest is to find a case in which a late-time accelerated expansion
is realized by the presence of the GB term.
Such a possibility can be accommodated by using the de-Sitter fixed
point discussed in the previous section.
We would also like to make use of scaling solutions
during the matter-dominated era in order to alleviate
the coincident problem.

When $\lambda=\mu$ there exists a scaling matter era
with $0<\Omega_\phi, \Omega_{\GB}<1$, but
the system does not get away from the scaling regime
to give rise to a late-time acceleration.
We require that $\mu$ does not
equal to $\lambda$ in order to exit from the scaling
matter era.
Let us consider the approximate scaling solution
(\ref{scalmune}) which exists under the conditions
$x_3 \approx 0$ and $\sqrt{6}\lambda x_1/2=-\dot{H}/H^2$.
Perturbing Eq.~(\ref{auto3}) about the fixed point, we obtain
\bea
\frac{{\rm d}}{{\rm d}N} \delta x_3=
\frac{3(\mu-\lambda)}{Q+\lambda}
\delta x_3\,.
\eea
To get $w_{\rm eff} \simeq 0$ during the matter era,
we require $Q \ll \lambda$ from Eq.~(\ref{scalmune}).
Since we are now considering positive $\lambda$,
the system departs from $x_3=0$ for $\mu>\lambda$
whereas it approaches $x_3=0$ for $\mu<\lambda$.
As we mentioned in the previous section,
the stability of the perturbations $\delta x_{1}$ and $\delta x_2$
is ensured when the condition $-Q/(Q+\lambda) \le \Omega_{\phi}<1$
is satisfied, i.e., $\lambda (\lambda +Q)>3p_{,X} \ge -2Q(\lambda +Q)$.
For a non-phantom field ($p_{,X}>0$) with $Q=0$,
this condition reduces to $\lambda^2>3p_{,X}$.
{}From the above discussion we find the following
property for the stability of the scaling solution:
\begin{itemize}
\item (i) Saddle for $\mu>\lambda$ and $\lambda (\lambda +Q)>3p_{,X}
\ge -2Q(\lambda +Q)$.
\item (ii) Stable for $\mu<\lambda$ and $\lambda (\lambda +Q)>3p_{,X}
\ge -2Q(\lambda +Q)$.
\end{itemize}
Then in the case (i) the solutions can exit from the scaling regime
to connect to the dark energy era.

As we showed in the previous section, the de-Sitter point (F)
is stable for $\mu>\lambda$, whereas it is a saddle
point for $\mu<\lambda$.
Then it is clear that the scaling solution can
be connected to the de-Sitter solution provided
$\mu>\lambda$ and $\lambda (\lambda +Q)>3p_{,X}
\ge -2Q(\lambda +Q)$.
The de-Sitter solution is present only for a restricted class of
models. In what follows we shall investigate
cosmological evolution for such two models:
(i) an ordinary field and (ii) a tachyon field.

\subsection{An ordinary field with an exponential potential}

\begin{figure}
\includegraphics[height=3.0in,width=3.0in]{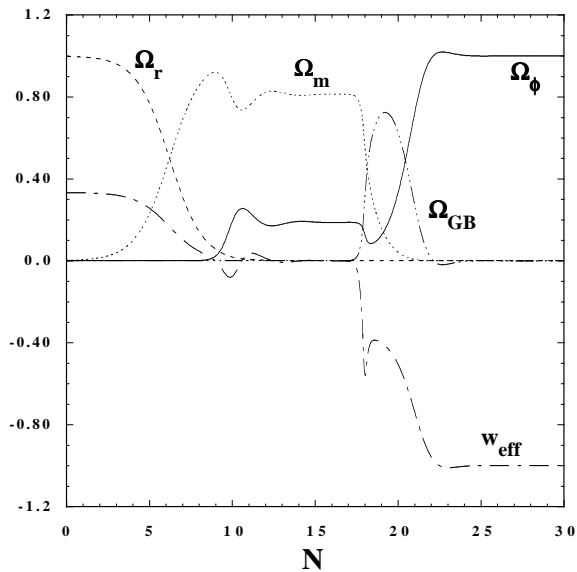}
\caption{Evolution of the variables $\Omega_{\phi}$, $\Omega_{\GB}$,
$\Omega_m$ and $\Omega_r$ together with
the effective equation of state $w_{\rm eff}$
for $\lambda=4$, $\mu=12$, $\alpha=10^{-22}$, $Q=0$
and $c=10^{-2}$ in the case of an ordinary field
with an exponential potential.
We choose initial conditions
$x_1=10^{-8}$, $x_{2}=10^{-7}$, $x_{3}=3.6 \times 10^{-9}$ and
$x_4=0.999$. The solution is in a scaling regime during the
matter-dominated epoch and finally approaches the de-Sitter
universe characterized
by $\Omega_{\phi}=1$, $\Omega_{\GB}=0$ and $w_{\rm eff}=-1$.
The energy fraction of the field $\phi$ during the scaling regime is
$\Omega_\phi =3/\lambda^2=0.1875$.}
\label{evo1}
\end{figure}

We shall first study the case of an ordinary field with an exponential
potential ($p=X-ce^{-\lambda \phi}$).
Since $p_{,X}=1$ in this case, one can have a scaling solution
in the matter era if $\lambda^2>3$ in the absence of the coupling $Q$.
In Fig.~\ref{evo1} we plot the evolution of $\Omega_{\phi}$,
$\Omega_{\GB}$, $\Omega_m$ and $\Omega_r$ together with
$w_{\rm eff}$ for $\lambda=4$, $\mu=12$, $\alpha=10^{-22}$
and $Q=0$.
The system starts from a radiation-dominated epoch,
which is followed by the scaling regime
in the matter-dominated era.
Since $\mu>\lambda$ the solution exits from the scaling regime and
finally approaches the stable de-Sitter fixed point.
The final attractor solution actually satisfies
$\Omega_{\phi}=1$ and $\Omega_{\rm GB}=0$,
as estimated analytically.

Figure \ref{evo1} shows that the energy fraction of the GB term 
grows right after end of the matter era, 
which begins to decrease after the increase
of $\Omega_{\phi}$. 
In Fig.~\ref{evo1} the effective equation of state
temporally takes a local minimum value 
$w_{\rm eff} \approx -0.6$ around 
$\Omega_m \approx 0.3$-$0.4$, which means that 
the accelerated expansion indeed occurs at the
present epoch in this scenario.
Note that the rapid transition of $w_{\rm eff}$ just after the
matter-dominated era is associated with the
growth of $\Omega_{\rm GB}$.
As $\mu$ is increased,  $w_{\rm eff}$ tends to get smaller.
When $\lambda=4$, for example, we find that the phantom
equation of state, $w_{\rm eff}<-1$, is realized for
$\mu>25$. Meanwhile the increase of $\mu$ leads to a shorter period
of the matter-dominated epoch.
Hence the transient phantom stage is realized at the expense
of such a short matter period.
If we take larger $\lambda$, it is also difficult to
get smaller values of $w_{\rm eff}$ satisfying the condition for
an accelerated expansion compatible with
observations. Thus $\lambda$ is bounded from above as well.

In Refs.~\cite{KM06,Koivisto06} Koivisto and Mota placed observational
constraints using the Gold data set of Supernova Ia \cite{Gold}
together with the CMB shift parameter data \cite{CMBshift}.
The parameter $\lambda$ is constrained to be 
$3.5 \lesssim \lambda \lesssim 4.5$ at the 90\%
confidence level, see Fig.~3 in Ref.~\cite{KM06}\footnote{
Note that the definition of $\lambda$ in Ref.~\cite{KM06}
is different from ours by the factor $\sqrt{2}$.}.
If the solutions are in the scaling regime
in radiation era the constraint coming from Nucleosynthesis gives
$\lambda>4.47$ \cite{Nelson,CST}, in which case the allowed range
of $\lambda$ is severely constrained.

\begin{figure}
\includegraphics[height=3.0in,width=3.0in]{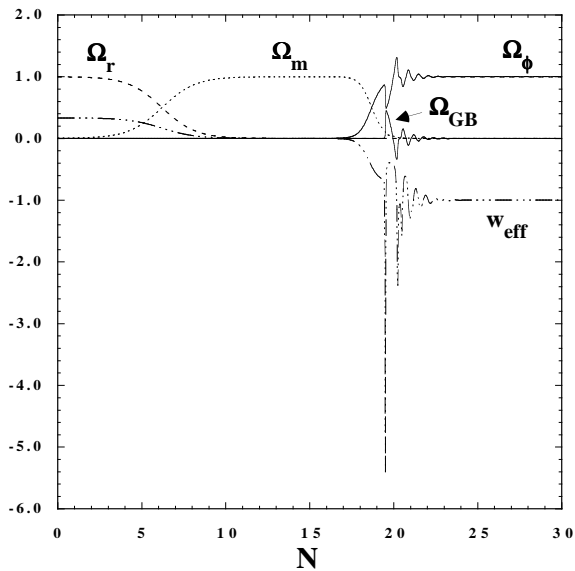}
\caption{Evolution of the same variables as in Fig.~\ref{evo1}
but for $\lambda=1$, $\mu=10^2$, $\alpha=10^{-29}$, $Q=0$
and $c=10^{-2}$.
We choose initial conditions
$x_1=10^{-13}$, $x_{2}=10^{-13}$, $x_{3}=3.3 \times 10^{-4}$
and $x_4=0.999$. After a standard matter era without a scaling regime,
the system falls down to a de-Sitter fixed point before it
reaches a scalar-field dominated point.
The effective equation of state temporally drops down toward
the phantom region ($w_{\rm eff}<-1$).
}
\label{evo2}
\end{figure}

When $\lambda=\mu$ we have shown that there exists a
scalar-field \& GB dominated point with
$\Omega_{\phi}+\Omega_{\GB}=1$.
This solution can be also used for a late-time acceleration.
Note that in the absence of the GB term the existence of
the stable scalar-field dominated solution demands the
condition $\lambda^2<3$ \cite{CLW}, in which
case the scaling solution with $\Omega_{\phi}<1$
does not exist in the matter-dominated epoch.
To get a stable scalar-field dominated solution,
one has to use the $\phi$MDE solution (\ref{phiMDE})
in the preceding matter period
(the standard matter era corresponds to $Q=0$).
In this case $x_{2}$ and $x_3$ need to be very small
during a radiation era to have $\Omega_{\phi} \ll 1$ and
$\Omega_{\GB} \ll 1$, which restricts the coupling $\alpha$
very small through the relation $x_2^2 x_3=\alpha/3$.
Then, at the scalar-field \& GB dominated point,
$\Omega_{\GB}$ in Eq.~(\ref{OmeGB})
is negligibly small relative to $\Omega_{\phi}$, in which
case the effect of the GB term is not important around
the present epoch.
Thus, for small $\lambda$,
the $\phi$MDE is followed by the
scalar-field \& GB dominated point,
but the impact of the GB term  for the late-time
acceleration is not significant for $\lambda=\mu$.

Then what happens for $\mu>\lambda$ ?
In this case the GB term finally comes out to
lead to the de-Sitter expansion by giving
rise to a minimum of an effective potential.
If $\mu$ is not much different from $\lambda$, this occurs
at sufficiently late-times after the solutions approach
the scalar-field dominated point.
On the other hand, if $\mu \gg \lambda$, the system can
be trapped at the de-Sitter point before it reaches
the scalar-field dominated solution.
In Fig.~\ref{evo2} we plot the cosmological evolution
for $\lambda=1$, $\mu=100$, $\alpha=10^{-29}$ and $Q=0$.
In this case the scaling solution does not exist in the matter era
because of the smallness of $\lambda$ ($<\sqrt{3}$).
We find that the GB term becomes important around $\Omega_\phi=0.7$,
which leads to the rapid decrease of $w_{\rm eff}$ toward the phantom
region. The effective equation of state approaches $-1$ after the
field settles down at the potential minimum.
The effect of the GB term can be seen at present time in such a case,
but the absence of the scaling matter era is not attractive to alleviate
the coincidence problem.

\subsection{A tachyon field with an inverse square potential}

As we have already mentioned, the tachyon field also
possesses the de-Sitter fixed point.
Meanwhile, when $Q=0$, the scaling solution does not
exist if the background fluid corresponds to
a non-relativistic matter ($w_m=0$).
In fact, for the function $g(Y)$ given in Eq.~(\ref{mo3}),
Eq.~(\ref{scalmune}) shows that the scaling
solution satisfies
\bea
\label{tac1}
& &\Omega_{\phi}=
\frac{1}{(Q+\lambda)^2} \left[ Q(Q+\lambda)
+\frac{3c}{\sqrt{1-2Y}} \right]\,, \\
\label{tac2}
& &\frac{c}{Y} \sqrt{1-2Y}=\frac23 Q (Q+\lambda)\,.
\eea
When $Q=0$ one gets $Y=1/2$ from Eq.~(\ref{tac2}), which
then means that $\Omega_{\phi}$ diverges from Eq.~(\ref{tac1}).
More generally we have $\Omega_{\phi}=3(1+w_m)/
(\lambda^2 \sqrt{-w_{m}})$ for the background fluid with an
equation of state $w_m$ \cite{Ruth,CGST}, showing that
the scaling solution exists only for $w_{m}<0$.
Hence in the absence of the coupling $Q$
one can not have a successful scaling epoch followed by
a de-Sitter expansion.

On the other hand the scaling matter era can exist if
the coupling $Q$ is present.
As we already mentioned, we require $Q \ll \lambda$ to recover
the equation of state $w_{\rm eff} \approx 0$.
In this case we get the approximate relation
$\sqrt{1-2Y} \approx Q\lambda/3c$ from Eq.~(\ref{tac2}).
Then from Eq.~(\ref{tac1}) we find that $\Omega_{\phi}$
is approximately given by
\bea
\Omega_{\phi} \approx
\frac{9c^2+(Q\lambda)^2}{Q \lambda^3}\,.
\eea
For example, when $Q=0.1$, $\lambda=10$ and $c=1$,
one has $\Omega_{\phi}=0.1$ and $w_{\rm eff}=-0.0099$.
Note that the variable $A$ defined in Eq.~(\ref{Adef}) is
given by $A=(1-2Y)^{3/2}/c$, which is positive for
$c>0$ (corresponding to a positive potential).
Then the scaling solution is stable provided that
$-Q/(Q+\lambda) \le \Omega_{\phi}<1$ and $c>0$.
It becomes a saddle point if the GB coupling
is present with $\mu$ larger than $\lambda$.
In this case the scaling matter era is followed by
the de-Sitter solution.

\begin{figure}
\includegraphics[height=3.7in,width=3.5in]{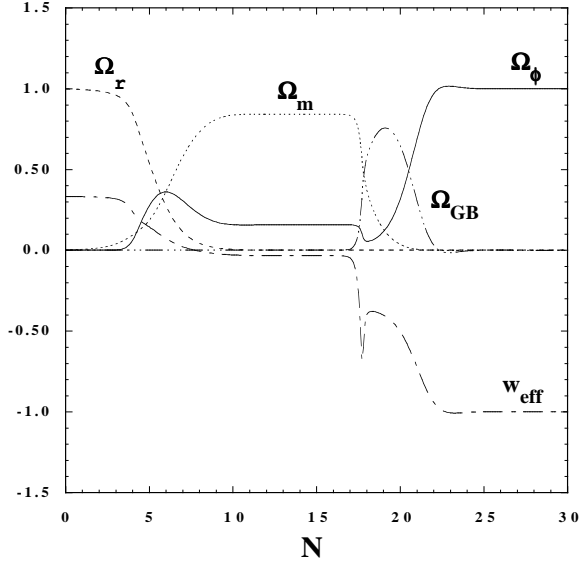}
\caption{Evolution of the variables $\Omega_{\phi}$, $\Omega_{\GB}$,
$\Omega_m$ and $\Omega_r$ together with
the effective equation of state $w_{\rm eff}$
for $Q=0.1$, $\lambda=7$, $\mu=20$, $\alpha=10^{-35}$
and $c=1$ in the case of a tachyon field with an
inverse square potential.
Initial conditions are chosen to be
$x_1=10^{-6}$, $x_{2}=10^{-4}$,
$x_{3}=3.0 \times 10^{-28}$ and $x_4=0.999$.
The scaling matter era is present because of the presence
of the coupling $Q$.}
\label{tachyon}
\end{figure}

In Fig.~\ref{tachyon} we plot the cosmological evolution
for $Q=0.1$, $\lambda=7$, $\mu=20$ and
$\alpha=10^{-35}$. We find that the solution first enters
the scaling matter era during which $\Omega_{m}$ and
$\Omega_{\phi}$ are constants.
This regime is followed by the growth of the GB term, which
leads to a rapid decrease of $w_{\rm eff}$ toward
$w_{\rm eff} \approx -0.7$ around the present time
($\Omega_m=0.3$-0.4). Numerically we checked that
the minimum value of $w_{\rm eff}$ temporarily reached
after the scaling matter era gets smaller if we choose smaller
$\lambda$ or larger $\mu$. Finally the solution falls down
to the de-Sitter fixed point with $\Omega_{\phi}=1$,
$\Omega_{\GB}=0$ and $w_{\rm eff}=-1$.

{}From the above discussion we require that $Q$ is bounded
from below to get the scaling matter era and that it is bounded
from above to obtain the effective equation of state
which does not differ from $w_{\rm eff}=0$ much.
It is certainly of interest to find the parameter spaces of $Q$
and $\lambda$ to realize the sequence of the scaling matter
era and the de-Sitter expansion satisfying observational
constraints, which we leave for future work.

\subsection{Ghost conditions}

Recently a number of authors \cite{Dede,Cal} discussed ghost conditions 
in GB cosmologies. It is possible to see the signature of ghosts
by studying gravitational perturbations.
We shall consider scalar and tensor perturbations about 
the FRW background \cite{mper}:
\begin{eqnarray}
\rd s^2 = - (1+2A)\rd t^2 +
2a\partial_iB \rd x^i\rd t
+a^2\left[ (1+2\psi)\delta_{ij}
+2\partial_{ij}E+2h_{ij}
\right] \rd x^i \rd x^j\,,
\end{eqnarray}
where $\partial_i$ represents the spatial partial derivative
$\partial/\partial x^i$. 
Here $A$, $B$, $\psi$ and $E$ are scalar quantities, whereas 
$h_{ij}$ is a tensor quantity.
We shall investigate whether or not ghosts appear
for an ordinary scalar field with an exponential potential,
i.e., $p=X-ce^{-\lambda \phi}$.
In what follows we neglect the contribution of the matter 
fluid. Although the equation of scalar perturbations is modified
in the presence of matter, the equation of tensor perturbations
is unchanged.

Defining the gauge-invariant comoving perturbation, 
${\cal R} \equiv \psi-H \delta{\phi}/\dot{\phi}$, 
the Fourier modes of scalar perturbations 
satisfy \cite{Hwang,GOT} 
\begin{eqnarray}
\frac{1}{a^3 Q_{\rm S}} (a^3 Q_{\rm S} \dot{\R})^\cdot
+c_{\rm S}^2 \frac{k^2}{a^2} {\R} =0\,,
\end{eqnarray}
where $k$ is a comoving wavenumber and
\begin{eqnarray}
& &Q_{\rm S} = \frac{(1-8H\dot{f}) \left[ (1-8H\dot{f})
\dot{\phi}^2+96 (H^2 \dot{f})^2 \right]}
{H^2 (1-12H\dot{f})^2}\,, \\
& & c_{\rm S}^2=\frac{(1-8H\dot{f}) \left[ (1-8H\dot{f}) \dot{\phi}^2+
96(H^2 \dot{f} )^2 +128 H^2 \dot{H}\dot{f}^2  \right]+
256 (H^2 \dot{f})^2 (\ddot{f}-H\dot{f})}
{(1-8H\dot{f})  \left[ (1-8H\dot{f}) \dot{\phi}^2+
96(H^2 \dot{f} )^2 \right]}\,.
\end{eqnarray}
Decomposing tensor perturbations into eigenmodes of the
spatial Lagrangian, $\nabla^2 e_{ij}=-k^2e_{ij}$, with
scalar amplitude $h(t)$, i.e., $h_{ij}=h(t)e_{ij}$,
where $e_{ij}$ have two polarization states,
the Fourier modes of tensor perturbations
obey the equation of motion \cite{Hwang,GOT} 
\begin{eqnarray}
\frac{1}{a^3 Q_{\rm T}} (a^3 Q_{\rm T} \dot{h})^\cdot
+c_{\rm T}^2 \frac{k^2}{a^2} {h} =0\,,
\end{eqnarray}
where 
\begin{eqnarray}
Q_{\rm T} =1-8H\dot{f}\,,\quad
c_{\rm T}^2=\frac{1-8\ddot{f}}
{1-8H\dot{f}}\,.
\end{eqnarray}

The perturbations exhibit exponential instabilities if $c_{\rm S}^2$ and $c_{\rm T}^2$
are negative. The propagation speeds of scalar and tensor 
modes are superluminal if $c_{\rm S}^2$ and $c_{\rm T}^2$
are larger than unity. In Refs.~\cite{Dede,Cal} it was shown that 
the no-ghost state to ensure a consistent quantum field theory gives the 
constraints $1-8H\dot{f}>0$ and $1-8\ddot{f}>0$.
Thus stable, non-superluminal and no ghost states 
require the conditions
\begin{eqnarray}
0 \le c_{\rm S}^2 \le 1\,, \quad
0 \le c_{\rm T}^2 \le 1\,, \quad
\xi \equiv 1-8H\dot{f}>0\,.
\end{eqnarray}
%

\begin{figure}
\includegraphics[height=3.5in,width=3.5in]{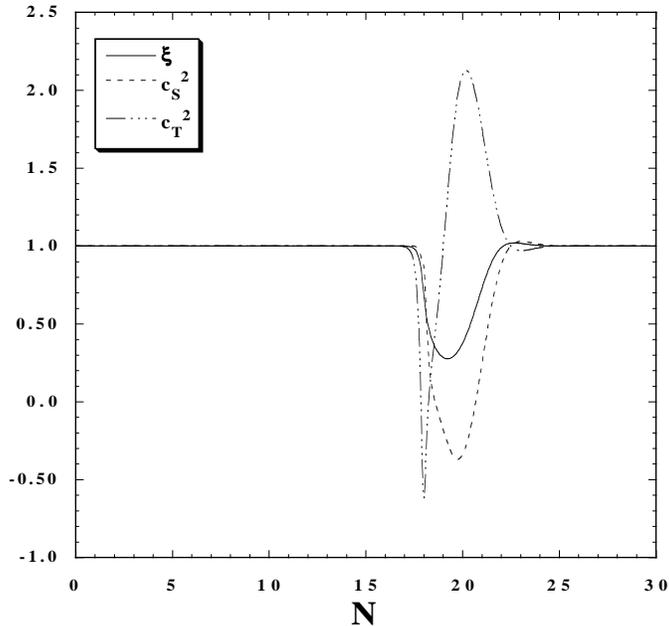}
\caption{Evolution of the variables $\xi=1-8H\dot{f}$,
$c_{\rm S}^2$ and $c_{\rm T}^2$
for an ordinary field with an exponential 
potential with the same model parameters
and initial conditions as given in Fig.~\ref{evo1}.
This shows that both $c_{\rm S}^2$ and $c_{\rm T}^2$
are temporally negative around the transition from the scaling 
matter era to the final de-Sitter era.
Moreover the propagation speed of the scalar mode
becomes superluminal.
}
\label{evo4}
\end{figure}

In Fig.~\ref{evo4} we plot the evolution of $\xi$,
$c_{\rm S}^2$ and $c_{\rm T}^2$ for the model 
$p=X-ce^{-\lambda \phi}$ with the same model 
parameters and initial conditions as given in Fig.~\ref{evo1}.
During the scaling matter era in which $x_3$ is much smaller
than of order unity [see Eq.~(\ref{x3ap})], we have that 
$|H \dot{f}| \ll 1$ and $|\ddot{f}| \ll 1$ giving 
$c_{\rm S}^2 \simeq 1$ and $c_{\rm T}^2 \simeq 1$.
For the de-Sitter point (\ref{desitter0}) one can easily 
show that $c_{\rm S}^2=c_{\rm T}^2=1$.
These properties are in fact confirmed in the numerical 
simulation of Fig.~\ref{evo4}.

We find from Fig.~\ref{evo4} that $c_{\rm S}^2$
and $c_{\rm T}^2$ temporally become negative
during the transition from the scaling matter era to 
the final de-Sitter era. This corresponds to the 
stage in which the contribution of the GB term 
is dominant. The perturbations exhibit exponential instabilities
associated with the appearance of ghosts.
The propagation speed of scalar perturbations 
can be modified in the presence of a matter fluid, but 
that of tensor perturbations is not affected.
Hence the negative instability of tensor perturbations
is robust even taking into account the contribution 
of the matter. This is consistent with the results of 
Ref.~\cite{GOT} in which the negative instability 
of tensor perturbations has been also found 
if inflationary solutions are constructed by 
the dominance of the GB term.
In Fig.~\ref{evo4} we also find that the propagation 
speed of scalar perturbations temporally 
become superluminal\footnote{In the context of K-essence, 
Bonvin {\it et al.} \cite{Bonvin} recently found that the propagation 
speed of scalar perturbations also temporally
becomes superluminal in order to solve the coincident problem 
of dark energy.}.

For the model parameters given in Figs.~\ref{evo1}
and \ref{evo4} the propagation speed 
$c_{\rm T}$ becomes imaginary 
before the system reaches the present epoch
(while $c_{\rm S}^2$ is positive).
Meanwhile if we choose smaller values of $\mu$, it is 
possible to avoid reaching the negative values of
$c_{\rm T}^2$.
For smaller $\mu$, however,  
the minimum value of $w_{\rm eff}$ tends to be larger.
Hence the exponential instability of tensor
perturbations can be avoided at the expense of losing 
rapid accelerated expansion of the universe.
It is of interest to find parameter ranges of $\lambda$
and $\mu$ which lead to positive $c_{\rm T}^2$
and also satisfy the observational constraints of
$w_{\rm eff}$.

\section{Conclusions}

In this paper we have discussed the viability of Gauss-Bonnet (GB)
dark energy models which possess cosmological scaling solutions. In
order to alleviate the coincident problem of dark energy, it is of
interest to find a cosmological trajectory which is in a scaling
regime ($\Omega_{\phi}/\Omega_{m}=$constant) during a matter era and
finally exits to give rise to a late-time accelerated expansion. We
have tried to understand the property of scalar-field dark energy
models which allow such a possibility.

First we derived the form of the GB coupling together with
the form of the scalar-field Lagrangian for the existence of
scaling solutions by starting from a very
general action given in Eq.~(\ref{eqn:action}).
The form of the GB coupling has been found
to be $f(\phi) \propto e^{\lambda \phi}$ together with
the field Lagrangian density $p=Xg(Xe^{\lambda \phi})$,
where $\lambda$ is defined
in Eq.~(\ref{lamdef}) and $g$ is an arbitrary function.
The field Lagrangian density takes the same form
as in the case where the GB coupling is absent.

We have also derived autonomous equations and fixed points
for the scaling Lagrangian $p=Xg(Xe^{\lambda \phi})$
with the GB coupling $f(\phi) \propto e^{\mu \phi}$.
When $\lambda=\mu$, we obtained the scaling solution
characterised by $x_1=\sqrt{6}/[2(Q+\lambda)]$ and
the scalar-field dominated solution characterised by
$\Omega_{\phi}+\Omega_{\GB}=1$ without specifying
any form of the Lagrangian.
In this case, however, the solutions do not get away from the
scaling matter regime to connect to a dark energy era.
When $\mu>\lambda$ the presence of the GB term can lead
to the exit from the scaling regime.

The GB term allows an interesting possibility to have a
late-time de-Sitter fixed point.
This point is present if the solutions for $gx_1^2=0$
and $x_1=0$ exist. The models which possess this de-Sitter
solution are characterised by the form $g(Y)=(c_0-c/Y)f(Y)$,
where $f(Y)$ is a function that approaches a constant
as $x_1  \to 0$. The ordinary field with an exponential potential
(\ref{mo1}) and the tachyon field with an inverse square
potential (\ref{mo3}) belong to this class,
whereas the dilatonic ghost condensate
model (\ref{mo2}) does not.

We have studied cosmological evolution for the
above two models in which the de-Sitter point is present.
In the case of an ordinary field with an exponential potential
one can in fact obtain a viable trajectory along which
the scaling matter period is followed by the de-Sitter solution
provided that $\mu>\lambda$ and $\lambda^2>3$
for $Q=0$. There is also an interesting possibility in which
the effective equation of state $w_{\rm eff}$ temporarily becomes
smaller than $-1$ because of the growth of the GB term.
This is realized by choosing larger values of $\mu$,
but it also corresponds to a shorter matter-dominated period.
The late-time acceleration is possible even for $\lambda=\mu$
by using a scalar-field \& GB dominated point ($\Omega_{\phi}+
\Omega_{\GB}=1$), but in this case the contribution of the
GB term is very small at the present epoch.
Moreover we do not have a scaling matter era in this case,
which is not a welcome feature in solving the coincidence problem.

The tachyon field has a de-Sitter fixed point, but in this case
the scaling solution is absent for a non-relativistic background
fluid ($w_{m}=0$) if $Q=0$.
Hence we do not have a successful sequence
of the scaling matter era and the de-Sitter expansion.
If the coupling $Q$ is present, however, it is possible to realize
this sequence when $\mu$ is larger than $\lambda$.
The values of $Q$ are bounded from both above and below
to get a scaling matter era with $w_{\rm eff}$ close to 0.
An alternative approach to the problem
discussed here can be provided by Renormalization Group (RG)
equations. The scaling solution presented here turns out to be
the fixed point of RG flow equations; the form of the field
Lagrangian and the GB coupling
are automatically fixed in this method \cite{sanjay}.

We also discussed ghost conditions for the GB dark energy 
models by considering gravitational perturbations.
For the ordinary field with an exponential potential we found that 
the propagation speeds of scalar and tensor modes can be 
imaginary during the transition from the scaling matter era to 
the final de-Sitter attractor. This is associated with the 
appearance of ghosts which lead to negative instabilities of 
perturbations. This ghost stage can be avoided by tuning the model 
parameters at the expense of obtaining a minimum effective 
equation of state closer to $-1$. 

In this paper we have not studied local gravity constraints on the
GB coupling \cite{ACD,ACD2}. If the contribution of the GB coupling
is large at present epoch, this may contradict with local gravity
experiments unless an accidental cancellation occurs to avoid the
large variation of effective gravitational constant. We leave future
work for the construction of scaling GB dark energy models
consistent with such constraints. We hope that the results obtained
in this paper will be useful to find such viable models.

\section*{ACKNOWLEDGEMENTS}
We thank L. Amendola, S.~Jhingan, S.~Nojiri
and S.~D.~Odintsov for useful discussions.
M.\,S. is supported by ICTP
and IUCAA through their associateship programs.
S.\,T. is supported by JSPS (Grant No.30318802).


\end{document}